\documentclass{aa}
\usepackage{psfig,times}
\usepackage{amssymb}
\usepackage{graphicx}

\begin{document}
\headnote{Research Note}
\title{Protoneutron star dynamos: pulsars, magnetars, and  
       radio-silent X-ray emitting neutron stars}

\author{
A.~Bonanno$^{1,2}$, V.~Urpin$^{1,3,4}$, and G.~Belvedere$^{1}$
}

\institute{
$^{1}$ INAF, Osservatorio Astrofisico di Catania,
        Via S.Sofia 78, 95123, Catania, Italy \\
        $^{2}$ INFN, Sezione di Catania, 
	Via S.Sofia 72, 95123, Catania, Italy \\
	$^{3}$ A.F. Ioffe Institute of Physics and Technology, 
        194021 St. Petersburg, Russia\\
        $^{4}$ Isaac Newton Institute of Chile, Branch in St. Petersburg,
        194021 St. Petersburg, Russia\\
}

\titlerunning{Protoneutron star dynamos}
\authorrunning{
Bonanno, Urpin, and Belvedere
}
\date{\today}

\abstract{We discuss the mean-field dynamo action in protoneutron stars 
          that are subject to instabilities during the early 
          evolutionary phase. The mean field is generated in the 
          neutron-finger unstable region where the Rossby number is 
          $\sim 1$ and mean-field dynamo is efficient. Depending on 
          the rotation 
          rate, the mean-field dynamo can lead to the formation of 
          three different types of pulsars. If the initial period of 
          the protoneutron star is short, then the generated large-scale 
          field is very strong ($> 3 \times 10^{13}$G) and exceeds the 
          small-scale field at the neutron star surface. If rotation is 
          moderate, then the pulsars are formed with more or less 
          standard dipole fields ($< 3 \times 10^{13}$G) but with 
          surface small-scale magnetic fields stronger than the dipole 
          field. If rotation is very slow, then the mean-field dynamo 
          does not operate, and the neutron star has no global field. 
          Nevertheless, strong small-scale fields are generated in such 
          pulsars, and they can manifest themselves as objects with 
          very low spin-down rate but with a strong magnetic field 
          inferred from the spectral features. 
 }

\maketitle

\section{Introduction}
A strong magnetic field in the neutron stars can be generated by 
turbulent dynamo during the first $\sim 30-40$s of their life when 
the  star is subject to hydrodynamic instabilities (Thompson \& 
Duncan 1993, Bonanno et al. 2003, 2005 (Paper I)). Instabilities in 
protoneutron stars (PNSs) are driven by either lepton or entropy 
gradients and, as a result, two different instabilities may occur 
(Bruenn \& Dineva 1996, Miralles et al. 2000). Convection is 
presumably connected to the entropy gradient, whereas the neutron-finger 
instability is caused by a lepton gradient. The neutron-finger 
instability grows on a timescale $\sim 30-100$ ms, that is 
$\sim 10-100$ times longer than the growth time of convection 
(Miralles et al. 2000). Turbulent motions generated by instabilities 
in combination with rotation make turbulent dynamo one of the most 
plausible mechanism of the pulsar magnetism. The character of 
turbulent dynamo depends on the Rossby number, $Ro = P/ \tau$, where 
$P$ is the PNS period and $\tau$ the turnover time. If $Ro \gg 1$, 
the effect of rotation on turbulence is weak and the mean-field 
dynamo is inefficient. If $Ro \leq 1$ and turbulence is 
strongly modified by rotation, the PNS can be subject to the mean-field 
dynamo action. The Rossby number is $\sim 10-100$ in the convective 
zone where the mean-field dynamo is not to work. On the contrary, 
except very slowly rotating PNSs, $Ro \sim 1$ in the neutron-finger 
unstable region (Bonanno et al. 2003) that favors the efficiency of 
the mean-field dynamo. Note that strong small-scale magnetic fields 
is generated by turbulence in the both unstable regions. 

The conclusion regarding the mean-field dynamo action is at variance 
with the results by Thompson \& Duncan (1993) 
who argued that only small-scale dynamos operates in most PNSs. 
Their reasoning is based on disregarding the neutron-finger 
instability and the assumption that the whole PNS is convective 
with the turbulent velocity $v_{T} \sim 10^{8}$ cm/s. Since 
$Ro \gg 1$ for convection, Duncan \& Thompson (1992) and Thompson \& 
Duncan (1993) concluded that the mean-field dynamo does not operate in 
PNSs, except those with $P \sim 1$ ms. Turbulent dynamo in PNSs has 
also been considered by Rheinhardt \& Geppert (2005) who modeled 
turbulent motions in the convective zone by a relatively complex but 
{\it stationary} velocity field. In reality, however, a velocity 
pattern changes on a timescale comparable to $\tau$ which 
is shorter than the growth time of a mean field. Therefore, the 
approach used is incorrect in principle because the stochastic nature 
of turbulent dynamo is entirely lost. In fact, the authors considered 
laminar rather than turbulent dynamo and, as a result, obtained a 
confusing result that a large-scale field can be generated in the 
convective zone where $Ro \gg 1$ in contradiction to the results by 
many authors for different astrophysical bodies (see, e.g., Thompson 
\& Duncan 1993, Chabrier \& K\"{u}ker 2005). 

If the magnetic field in PNSs is generated by turbulent dynamos one 
can then expect that the field of pulsars is complex with both 
small- and large-scale structures presented. After the unstable phase, 
small-scale fields decay rapidly due to ohmic dissipation but fields 
with the lengthscale $\gtrsim 1$ km can survive during the lifetime 
of radiopulsars (Urpin \& Gil 2004). 

In this note, we show that the PNS dynamo can consistently account 
for the origin of both ``standard'' radiopulsar magnetic fields and 
ultra-strong fields in ``magnetar''. Our model predicts also the 
existence of X-ray emitting isolated neutron stars which possess a 
strong small-scale field at their surfaces but exhibit no 
radio-pulsations because of a weak large-scale field.

\section{The model and results}
The problem for dynamo modelling has been described in Paper I. 
We model the PNS as a sphere of radius $R$ with two different 
turbulent zones separated at $R_{c}$. The inner part ($r<R_{c}$) 
corresponds to the convective zone, while the outer one ($R_{c}<r<R$) 
to the neutron-finger unstable zone. Generation is governed by the 
standard dynamo equation with the $\alpha$-term and turbulent magnetic 
diffusivity $\eta$ included. We consider three models of cylindrical 
rotation given by Eq.~(3) of Paper I with $\Omega_{cyl}^{(1)}=0, -1/2, 
-2/3$ (model a, b, c). 
Properties of turbulence are different at $0 \leq r \leq R_{c}$ and 
$R_{c} \leq r \leq R$. To model this, we assume that $\eta$ in the 
convective and neutron-finger unstable regions is equal to $\eta_c$
and $\eta_{nf}$, respectively. The $\alpha$-parameter is small in the 
convective zone and equal to $\alpha_{nf}=$const in the neutron-finger 
unstable region. We recall that in turbulence with the length-scale 
$\ell_{T}$ and moderate $Ro$, $\alpha_{nf} \approx \Omega 
\ell_{T}^{2} \nabla \ln (\rho v_{_{T}}^{2})$ (R\"{u}diger \& Kitchatinov 
1993). Denoting the density length-scale as $L$, we have $\alpha_{nf} 
\approx 2 \pi \varepsilon L/P$ where $\varepsilon = \ell_{T}^{2}/L^{2}$. 
Since the maximal length-scale of instability is $\sim L$, we have 
$\varepsilon \sim 1$. By solving the dynamo equation, we determine the 
critical period $P_{c} = \varepsilon P_{0}$ corresponding to the 
marginal dynamo stability; $P_{0}$ is the critical period for 
$\varepsilon =1$. Generation is possible if $P < P_{c}$ but PNSs with 
$P > P_{c}$ will not be subject to the mean-field dynamo action. The 
critical period is rather long ($P_{0} \sim 1$ s) and, likely, the 
mean-field dynamo is effective in most PNSs.
\begin{table}
\caption[]{
The critical periods $P_0$ for 
different PNS models}
\begin{center}
\begin{tabular}{l l l l l  }
\cline{1-4}
\noalign{\smallskip}
$P_0$ [s]  & $\eta_{nc}/\eta_{c}$ & model & parity  \\
\noalign{\smallskip}
\cline{1-4}
\noalign{\smallskip}
0.96 & 0.1    &  a & A0 &  \\
1.13 & 0.1    &  b & A0 &  \\
1.22 & 0.1    &  c & A0 &  \\
0.94 & 0.1    &  a & A1 &  \\
0.93 & 0.1    &  b & A1 &  \\
0.89 & 0.1    &  c & A1 &  \\
0.84 & 0.02   &  a & A0 &  \\
0.97 & 0.02   &  b & A0 &  \\
1.15 & 0.02   &  c & A0 &  \\
0.81 & 0.02   &  a & A1 &  \\
0.83 & 0.02   &  b & A1 &  \\
0.81 & 0.02   &  c & A1 &  \\
\noalign{\smallskip}
\cline{1-4}
\end{tabular}
\end{center}
\label{lista}
\end{table}

In Table 1, we compare the value of $P_0$ for PNS models with 
different $\eta_{c}$ and rotation laws. In calculations, we assume
$R_c/R=0.6$ and $\eta_{nf}= 10^{11}$ cm$^{2}$s$^{-1}$. The dependence 
of $P_0$ on the rotation law, $\eta_{c}$, and azimuthal wave number 
$m$ is rather weak. Clearly, larger $\eta_{c}$ enhances diffusion of 
the magnetic field from the neutron-finger unstable zone and, as a 
result, decreases the critical period. However, this decrease is not 
large, and $P_{0} \sim 1$ s for all considered models. 

\begin{figure}
\begin{center}
\includegraphics[width=8.0cm]{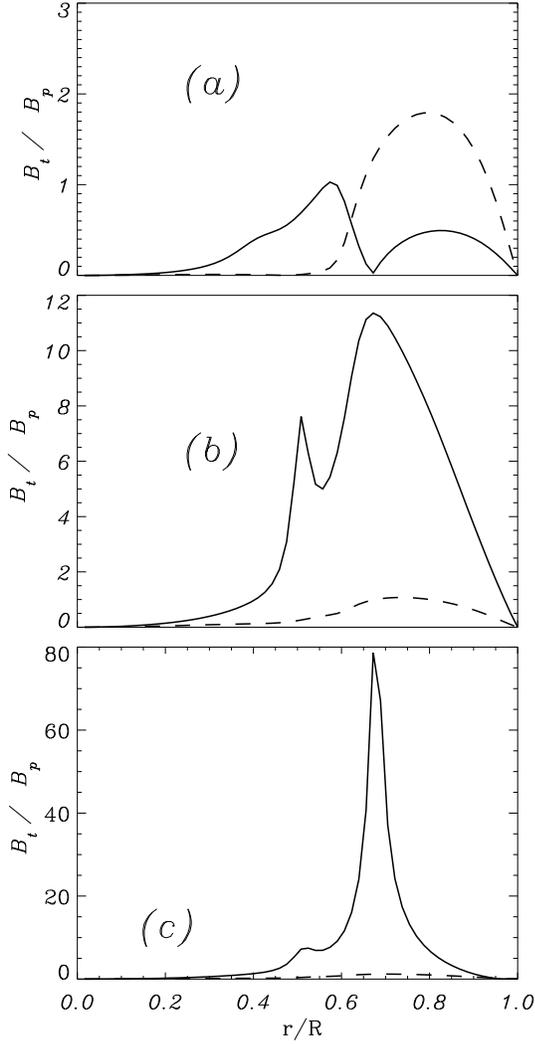}
\caption{The radial dependence of $B_{t}/B_{p}$ for $\varepsilon = 1$,
$\eta_{nf}/ \eta_{c} = 0.1$, and for the polar angle $\theta=60^{\circ}$ 
(solid lines) and $20^{\circ}$ (dashed lines).}
\end{center}
\end{figure}
  
\begin{figure}
\begin{center}
\includegraphics[width=8.0cm]{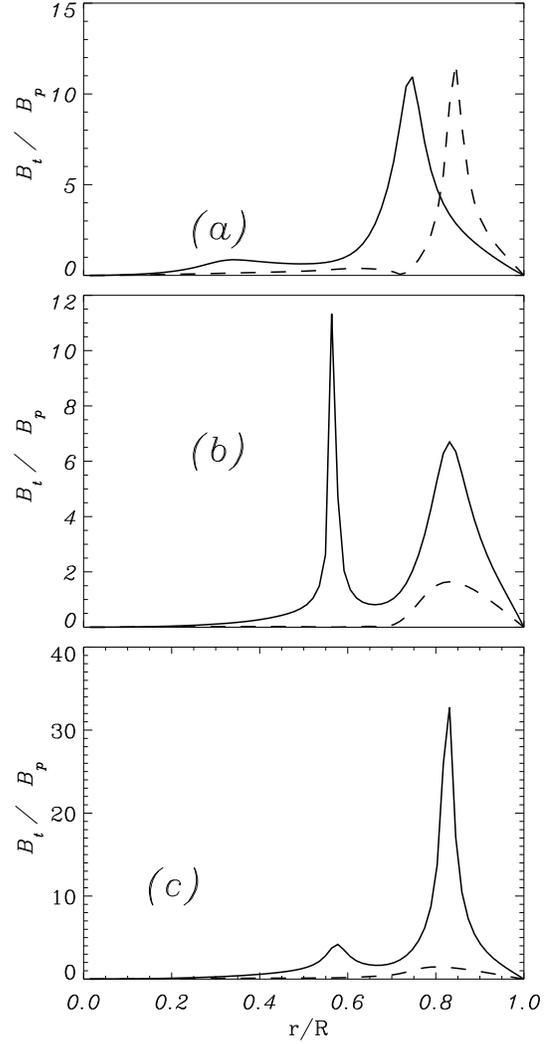}
\caption{Same as in Fig.~1 but for $\eta_{nf}/ \eta_{c} = 0.02$.}
\end{center}
\end{figure}
  
The geometry of the generated field is rather complex. The field 
is basically concentrated in the neutron-finger unstable region where 
the mean-field dynamo operates. The field in the convective zone is 
substantially weaker. In Fig.1, we plot the radial dependence of the 
ratio of toroidal $B_{t}$ and poloidal $B_{p}$ fields for the 
axisymmetric mode ($m=0$) and different rotation laws. If rotation 
is almost rigid, then the $\alpha^{2}$-dynamo is more efficient, and 
$\xi \equiv |B_{t}/B_{p}| \sim 1-3$ in the neutron-finger unstable 
region. However, $\xi$ is larger if differential rotation 
is strong. For instance, $\xi \sim 6-10$ and $\sim 20-60$ for our 
models (b) and (c). In these cases, the generation is determined by 
a combined effect of the $\alpha^2$- and $\alpha \Omega$-dynamo, and 
the toroidal field is noticeably stronger. The field tends to 
concentrate near the polar region for the $\alpha^2$-dynamo and 
closer to the equator for the $\alpha \Omega$-dynamo. The values of 
$\xi$ are approximately same for the $m=1$ mode. In Fig.~2, we show 
same as in Fig.~1 but for the model with $\eta_{nf}/\eta_{c} =0.02$. 
More efficient diffusion into the convective region decreases the 
ratio $\xi$, particularly, for the models (b) and (c) but this 
decrease is not significant and does not change our main conclusions. 
  
Generally, the steady-state $\Omega$-profile in turbulent stars 
depends on $P$, as it emerges both from observations and numerical 
modelling. For instance, Reiners \& Schmitt (2003) found that 
differential rotation in convective F-stars is much more common for 
stars with moderate or slow rotation than for rapid rotators. 
Simulations also show that the $\Omega$-profile depends on 
$Ro$ in convective stars (see, e.g., K\"uker \& R\"udiger 2005). 
Rotation is almost rigid in fast rotators ($Ro \leq 1$) whereas 
slow rotators ($Ro \gg 1$) may exhibit a significant differential 
rotation. Therefore, we can expect that PNSs with $Ro \leq 1$ in the 
neutron-finger unstable zone (or with $P < 30-100$ ms) rotate 
approximately rigidly and are subject to the $\alpha^{2}$-dynamo. 
For such stars, $B_{t} \sim B_{p}$ and the internal and surface 
mean fields are comparable. Slowly rotating PNSs ($P > 30-100$ ms) 
can be differential rotators. Likely, the $\alpha^{2}$-dynamo is 
accompanied by the $\alpha \Omega$-dynamo in such stars, and the 
internal toroidal field can be substantially stronger than the 
surface field.  
 
\section{Discussion}

The critical period that determines the onset of mean-field dynamo 
is rather long, and dynamos should be effective in most PNSs. The 
unstable stage lasts $\sim 40$ s, and this is sufficient for dynamo 
to reach saturation. We can estimate a saturation field assuming the 
simplest $\alpha$-quenching with non-linear $\alpha$ given by 
$\alpha (\tilde{B}) = \alpha_{nf} (1 + \tilde{B}^{2}/B_{eq}^{2})^{-1}$,  
where $\tilde{B}$ is a characteristic strength of the generated field,
and $B_{eq}$ is the equipartition small-scale field. The saturation is 
reached when $\alpha$ becomes equal to $\alpha_{c}$ corresponding to 
the marginal stability. This yields
\begin{equation}
B_{s} \approx B_{eq} \sqrt{P_{c}/P-1} .
\end{equation}
A further activity of the PNS as a radiopulsar is determined by the 
poloidal component of this field, $B_{ps}$. Using the estimate 
$B_{s} \sim B_{ps} (1 + \xi)$, we obtain 
\begin{equation}  
B_{ps} \approx B_{eq} (1 + \xi)^{-1} \; \sqrt{P_{c}/P-1},
\end{equation}
The equipartition field, $B_{eq} \approx 4 \pi \rho v_{T}^{2}$, varies 
during the unstable stage, rising rapidly soon after collapse and then 
going down when the temperature and lepton gradients are smoothed. We 
can estimate $B_{eq}$ at a peak as $\sim 10^{16}$ G in the convective 
zone and $\sim (1-3) \times 10^{14}$ G in the neutron-finger unstable 
zone (Urpin \& Gil 2004). However, $v_{T}$ and $B_{eq}$ decrease 
whereas the turnover time $\tau$ increases as the PNS cools down. 
The meand-field dynamo description applies as the quasi-steady 
condition $\tau_{cool} \gg \tau$ is fulfilled with $\tau_{cool}$ being 
the cooling timescale. We assume that the final strength of the 
generated magnetic field is determined by $B_{eq}$ at the moment when 
the quasi-steady condition breaks down and $\tau \sim \tau_{cool}$. 
This is in a contrast to the assumption made by Thompson \& Duncan 
(1993) that the final field strength is approximately equal to the 
maximal one. If $\tau \sim \tau_{cool}$, then we have $v_{T} \sim \pi 
\ell_{T} /\tau_{cool}$ and
\begin{equation}
B_{eq} \sim  \sqrt{4 \pi \rho} v_{T} \sim 
\pi \sqrt{4 \pi \rho} \ell_{T} \tau_{cool}^{-1}.
\end{equation}
The final equipartition field is same for the both unstable zones.
Estimate (3) yields $B_{eq} \sim (1-3) \times 10^{13}$ G for $\ell_{T} 
\sim 1-3$ km if $\tau_{cool} \sim$ few seconds. We can distinguish 
three types of neutron stars which exhibit different magnetic 
characteristics. 

{\it Strongly magnetized neutron stars.} If the period satisfies 
the condition $P < P_{m} \equiv P_{c} [1 + (1 + \xi)^{2}]^{-1}$ then 
dynamo leads to the formation of a strongly magnetized PNS with $B_{ps} 
> B_{eq} \sim 3 \times 10^{13}$G. Since $Ro \leq 1$ for fast rotators, 
we can expect that such stars rotate almost rigidly, and the 
$\alpha^{2}$-dynamo is operative. Therefore, $\xi \sim 1-3$ in the 
neutron-finger unstable region, and such strongly magnetized stars can 
be formed if $P$ is shorter than $\sim 0.1 P_{c} = 0.1 \varepsilon 
P_{0}$. Then, Eq.~(2) yields 
\begin{equation}
B_{ps} \sim 0.3 B_{eq} \sqrt{P_{c}/P -1 } \sim 10^{13} 
\sqrt{P_{c}/P -1} \;\;\; {\rm G}.
\end{equation}   
The shortest possible period is $\sim 1$ ms, hence, the strongest 
magnetic field generated by dynamo is $\sim 3 \times 10^{14}$ 
G. This is a bit higher than the maximum field inferred from 
the spin-down data in radiopulsars, $\sim 9.4 \times 10^{13}$ G 
(McLaughlin et al. 2003). Likely, the field of all five 
high-magnetic-field radiopulsars known at the moment such as
PSR J1847-0130 ($9.4 \times 10^{13}$ G), PSR J1718-3718 ($7.4 \times 
10^{13}$ G), PSR J1814-1744 ($5.5 \times 10^{13}$ G), PSR J1119-6127 
($4.4 \times 10^{13}$ G), and PSR B0154+61 ($2.1 \times 10^{13}$ G) 
has been generated in this regime, and these pulsars had $P \sim$few 
ms at their birth. Note that some PNSs which were strongly magnetized 
just after the generation stage, now may possess a dipole field $< 3 
\times 10^{13}$ G since a fraction of the electric current decays 
during the early evolution when the conductivity is relatively low 
(Urpin \& Gil 2004). Since the small-scale field is weaker than the 
large-scale field, one can expect that radiopulsations from such stars 
may have a more regular structure than those from low-field pulsars.

The maximum field generated is comparable to estimates of the surface 
field in some SGRs and AXPs which are believed to be the candidates 
in magnetars. However, estimates of $B$ inferred from spin-down are 
not reliable for AXPs and SGRs because $\dot{P}$ can vary significantly 
on a short timescale that seems to be unrealistic 
for the magneto-dipole braking (Kaspi \& McLaughlin 2004).   

{\it Neutron stars with moderate magnetic fields.} If $P_{c} > P > 
P_{m}$ (or. approximately, $\varepsilon P_{0} > P > 0.1 \varepsilon
P_{0}$) then the generated dipolar field is weaker than the small-scale 
field, $B_{ps} < B_{eq} \sim 3 \times 10^{13}$G. Since $Ro \sim 1$ or 
slightly larger for such PNSs, their rotation can be differential. 
Departures from rigid rotation are weak if $P \sim P_{m}$ but can be 
noticeable for $P \sim P_{c}$. As a result, $\xi$ can vary within 
a wide range for these PNSs, $\xi \sim 5-60$, and $B_{t} \gg B_{p}$. 
The toroidal field is stronger than the small-scale field if $P_{c}/2 
> P > P_{m}$, and weaker if $P_{c} > P > P_{c}/2$. Note that both 
differential rotation and strong toroidal field can influence the 
thermal evolution of such stars. Heating caused by dissipation of 
differential rotation is important during the early cooling phase 
since viscosity operates on a timescale $\sim 10^{2}-10^{3}$ yrs. On 
the contrary, ohmic dissipation of the toroidal field is a slow 
process and can maintain the surface temperature $\sim (1-5) \times 
10^{5}$K during $\sim 10^{8}$ yrs (Miralles et al. 1998). 

The magnetic field of this group of stars should be very irregular 
with a number of sunspot-like magnetic structures on their surface. 
Particularly, this concerns slowly rotating PNSs with $P \sim P_{c}$ 
where only a weak dipole field can be generated. Urpin \& Gil (2004)
argued that magnetic spots with the lengthscale $\gtrsim 1$ km can 
survive during the lifetime of radiopulsars. Therefore, one can 
expect that radiopulsations from these pulsars may have a complex 
behaviour. For example, small-scale magnetic structures can be 
responsible for the phenomenon of drifting subpulses observed in many 
pulsars (Gill \& Sendyk 2003). Note also that features in the X-ray 
spectra of these pulsars may indicate the magnetic field which 
differs essentially from that inferred from the spin-down data. 
This can happen because spectral features provide information about 
the strength of small-scale fields at the surface rather than about 
the mean field associated to the magneto-dipole braking.  

{\it Neutron stars with no large-scale field.} If $P$ is longer than 
$P_{c}= \varepsilon P_{0}$ then the mean-field dynamo does not 
operate but the small-scale dynamo can still be efficient. We
expect that such neutron stars have only small-scale fields with the 
strength $B_{eq} \sim 3 \times 10^{13}$ G and no dipole field. If 
$\varepsilon \sim 0.3$ then such object can be formed if the initial 
PNS period is $> 0.3$ s. Likely, such slow rotation is rather 
difficult to achieve, and the number of such exotic PNSs is small. 

The characteristics of pulsars originated from such PNSs should
be rather unexpected. Since the dipole field is negligible, the
spin-down rate should be low for these objects (if spin-down is 
determined by the magneto-dipole braking alone). Therefore, 
periods of such pulsars do not increase much in the course of evolution. 
These neutron stars cannot manifest themselves as bright radiopulsars 
because of a negligible dipole field. Most likely, they are 
almost radio-silent. Nevertheless, they can be observed due to 
periodicity in X-rays caused by the presence of small-scale magnetic 
structures at their surface. Pulsations must be weak because 
inhomogeneities of the surface temperature caused by a nonuniform 
magnetic field $\sim 3 \times 10^{13}$ G are rather small (e.g., Potekhin 
\& Yakovlev 2001). Note that the thermal evolution of these objects 
should be very similar to that of pulsars with a moderate magnetic 
field, and they can be X-ray emitting during a rather long time 
$\sim 1$ Myr. Likely, the most remarkable property of these neutron 
stars is a discrepancy between the magnetic field that can be 
inferred from spin-down measurements and the field strength obtained 
from spectral observations. Features in X-ray spectra may indicate the 
presence of a rather strong magnetic field $\sim 3 \times 10^{13}$ G 
(or a bit weaker because of ohmic decay during the early evolution) 
associated to sunspot-like magnetic structures at the surface. The field 
inferred from  spin-down data should be essentially lower.

\noindent
{\it Acknowledgements.} VU thanks {\it Dipartimento di Fisica ed 
Astronomia}, University of Catania, for financial support.

{}

\end{document}